\documentclass[reprint,amsmath,amssymb,aps,showpacs,pra,floatfix]{revtex4-1}
\usepackage{graphicx}
\usepackage{dcolumn}
\usepackage{bm}
\def\av#1{\langle#1\rangle}
\renewcommand{\l}{\ell}
\newcommand{\ls}{\ell_{\textrm{s}}}
\newcommand{\lu}{\ell_{\textrm{u}}}

\newcommand{\la}{\langle}
\newcommand{\ra}{\rangle}
\newcommand{\be}{\begin{equation}}
\newcommand{\ee}{\end{equation}}

\begin{document}
\title[Path sampling]{Spreading paths in partially observed social networks}
\author{Jukka-Pekka Onnela$^{1}$}
\email{Onnela@med.harvard.edu}
\homepage{jponnela.com}
\author{Nicholas A. Christakis$^{1,2,3}$}
\affiliation{
$^1$\mbox{Department of Health Care Policy, Harvard Medical School, Boston, MA, USA}\\
$^2$\mbox{Department of Medicine, Harvard Medical School, Boston, MA, USA}\\
$^3$\mbox{Department of Sociology, Harvard Faculty of Arts and Sciences, Cambridge, MA, USA}}

\begin{abstract}
Understanding how and how far information, behaviors, or pathogens spread in  
social networks is an important problem, having implications for both  
predicting the size of epidemics, as well as for planning effective  
interventions. There are, however, two main challenges for inferring  
spreading paths in real-world networks. One is the practical difficulty  
of observing a dynamic process on a network, and the other is the  
typical constraint of only partially observing a network. 
Using a static, structurally realistic social network as a platform for  
simulations, we juxtapose three distinct paths: (1) the stochastic  
path taken by a simulated spreading process from source to target; (2) 
the topologically shortest path in the fully observed network, and hence the single most likely  
stochastic path, between the two nodes; and (3) the topologically shortest path 
in a partially observed network. In a sampled network, how closely  
does the partially observed shortest path (3) emulate the unobserved  
spreading path (1)? Although partial observation inflates the length of  
the shortest path, the stochastic nature of the spreading process also  
frequently derails the dynamic path from the shortest path. We find that the  
partially observed shortest path does not necessarily give an inflated estimate of the  
length of the process path; in fact, partial observation may,  
counterintuitively, make the path seem shorter than it actually is.
\end{abstract}

\pacs{89.75.-k, 89.75.Hc, 02.50.Tt}

\maketitle

\section{Introduction} \label{sec:intro}
The small-world property, first empirically discovered by Milgram \cite{milgram} and then revisited by many, perhaps most famously by Watts and Strogatz \cite{smallworld}, captures the remarkable idea that we are all connected to each other via very short paths, typically encompassing only a handful of intermediaries. Path-based network measures, such as diameter and average path length, are useful elementary network characteristics, but exploring paths and path lengths is especially important when dealing with processes on networks that may be able permeate only up to a finite depth. This is relevant for a large class of general infection processes, such as the propagation of a certain behavior, the transmission of a piece of information, or the spread of a pathogen. As a first approximation, one might, of course, assume that any of these may percolate through entire social networks, and indeed the relationship between the three paths discussed in this paper also hold in that case. However, it is likely in practice that the information being transmitted gets altered along the way;  the behavior gets modified as it is imitated; or the pathogen becomes mutated as it is passed on. Consequently, the penetration depth of a given piece of information, a given behavior, or a given pathogen is often bounded. When this is the case, understanding path lengths becomes especially important, 

More nuanced accounts of spreading phenomena should distinguish between these different variants of information, behaviors, or pathogens. When viewed from this angle, any given spreading processes in itself, most likely, has a finite  (typically stochastic) permeation depth. But measuring these depths is difficult in practice because of the fundamental difficulty in monitoring the unfolding of real-world spreading processes. Even when time-stamped interaction events are available, such as in some recent insightful studies utilizing cell phone communication data \cite{karsai2011,kumar2011}, one still does not have actual spreading data but, instead, needs to assume that something is being spread, possibly across multiple ties, and one also needs to operationalize this assumption. (We would like to point out to the reader that the notion of temporal distance, corresponding to the time-ordered shortest path between nodes and defined for empirical event sequences in \cite{kumar2011}, is different from the notion of dynamic path lengths discussed below.) In contrast, unlike the process itself, outcomes of a spreading process are often directly observable (e.g. symptoms of chicken pox). But even if we could observe the outcomes, a key remaining challenge in dealing with person-to-person social networks is that instead of observing the full network evolve in time, financial and human resources, ethical considerations, and methodological issues typically limit us to a sampled, or partially observed, network snapshot (an exception is experimental networks \cite{centola2010,christakis_pnas_coop}).

Transport processes, such as the routing of data on the Internet \cite{vespignani2007}, are somewhat different from but related to the spreading processes discussed above. In contrast to the World Wide Web, which allows for the links from each site to be observed, it is not possible to directly map the physical connections between Internet routers. Instead, these networks are typically sampled using traceroute-like methods, which are trees initiated from a single source. It has been recently shown both empirically \cite{lakhina2003} and analytically \cite{clauset2005,achlioptas2005} that the resulting sampled networks are biased \footnote{For example, the underlying Poisson degree distribution of an ER graph may erroneously appear as a power law in the sample.}. Notwithstanding the common assumption that data packets follow shortest routes from source to target, it was found that, although the undirected shortest paths had a mean length of $11.4$, the routes had a mean length of $15.6$ hops \cite{leguay2004} and only $19.3$\% of the routes taken were along the shortest paths.

Here, we focus on the problem of estimating infection path lengths for an unobsevable stylized infection process in a partially observed social network. Similar to degrees of separation, which quantify how far nodes are from each other, infection path lengths, also known as degrees of influence, quantify how far a given process might spread in the network \cite{connected}. In the case of contact networks, understanding path lengths might enable us to estimate the virulence of a pathogen and its nature, e.g., how frequently it mutates. In the case of social networks, understanding path lengths might enable us to evaluate the infectiousness of certain behaviors and experiences, such as obesity, depression, voting, and smoking \cite{nac_obesity,nac_smoking}. Understanding how far these conditions may be able to spread from one person to another has important consequences for both gauging the overall extent of these ``social epidemics," as well as for planning the most effective interventions. Both goals are of substantial importance from the point of view of public policy.

Since one cannot in practice follow the paths taken by an actual infection or spreading process, the shortest path connecting the source and target nodes functions as a reasonable proxy for the actual path. Indeed, the shortest path is the single most likely path connecting a given source node to a given target node, since the probability for a given path, under some fairly general assumptions, decreases exponentially as a function of its length. A counterbalancing factor is that the number of paths, or path degeneracy, increases as a function of the distance between the source and target nodes, and this happens in a way that depends delicately on the structure of the network. An important consequence is that spreading phenomena often do not follow the shortest paths. Still, all in all, the shortest path is always our best guess for the actual path, given that in a practical setting one does not have microscopic spreading data available.

\begin{figure}
\begin{center}
\includegraphics[width=1\linewidth]{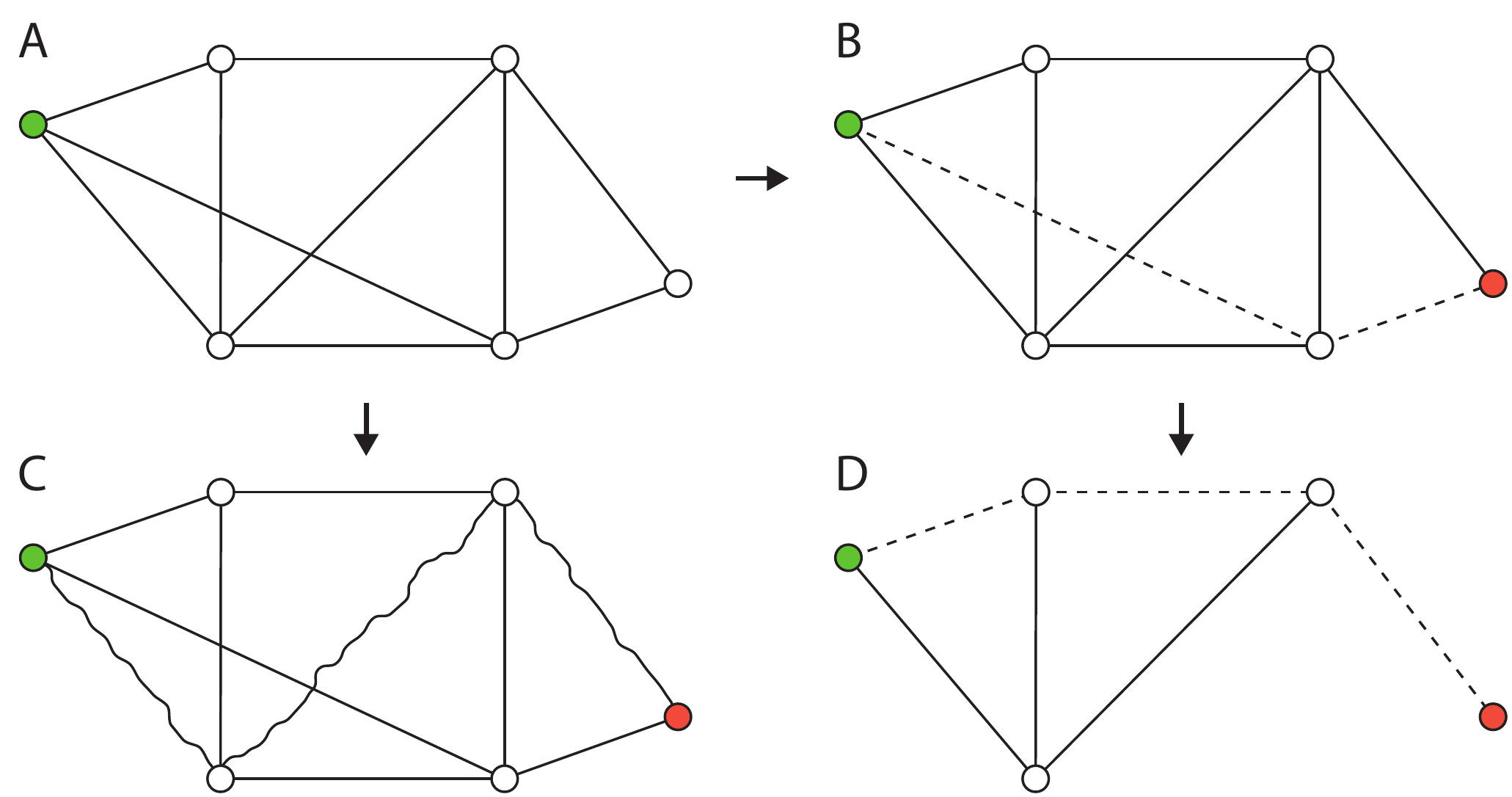}
\caption{\label{fig:schem} (Color online) 
Schematic of a network infection and sampling process.
(\textbf{A}) The full (unobserved) network with the initially infected node colored (upper left corner).
(\textbf{B}) The shortest path from the source node to the target node (lower right corner) corresponds to the most likely infection path in the fully observed network and has $\lu = 2$.
(\textbf{C}) The (unobservable) spreading process unfolds in the (unobserved) network. The actual path taken by the infection is shown with wavy edges. The target node is reached in three steps giving $\l = 3$. 
(\textbf{D}) The partially observed network has some nodes and links missing depending on the sampling parameters. The shortest path from source to target has length $\ls = 3$, corresponding to the length of the most likely path taken by the infection. In this case, using the shortest path length in the fully observed network $\lu$ to estimate the actual path length $l$ would result in an underestimate of path length, whereas using the path in the partially observed network, in this case, correctly yields $\l = \lu = 3$.}
\end{center}
\end{figure}

This results in three different paths to consider (Fig.~\ref{fig:schem}). First, there is the stochastic path of length $\l$ from node $i$ to node $j$, followed by the as-yet-unspecified but inherently unobservable dynamic process; second, there is the unsampled, potentially observable, but often only partially observed, shortest path of length $\lu$ (subscript $u$ for ``unsampled'') between nodes $i$ and $j$; and, finally, there is the shortest path in the sampled network of length $\ls$ (subscript $s$ for ``sampled'') from node $i$ to $j$.

As mentioned above, the relationship between the three paths holds whether or not the spreading process has a finite permeation depth. However, when this is the case, the problem becomes even more relevant, because now the properties of the thing that is spreading might be  related to the length of the actual path it has taken through the system. For example, the relative stability or mutability of pathogens can depend on the properties of the system through which they are moving. Recently, genotyping of pathogens has been combined with social network mapping to identify likely point sources of epidemics, and infection paths; this work has contrasted biological and social network approaches to identifying and quantifying outbreaks \cite{gardy2011}.

We will explore some of the properties of these three distinct paths by using a real-world social network as a platform for simulating both the spreading process and the subsequent sampling process. We introduce the dataset in Section \ref{sec:data}, and describe the details of our approach in Section \ref{sec:methods}. The main results are presented in Section \ref{sec:results}, and we discuss our findings Section \ref{sec:discussion}.

\section{Community networks} \label{sec:data}
In this Section, we study path lengths for a simple spreading process on a static real-world social network with sampling. The platform network possess all the prototypical features of social networks: a fat-tailed degree distribution, assortativity by degree, a high level of clustering, the small-world property, and network communities. Our results are therefore expected to hold for social (and other) networks with similar characteristics. The platform is a communication network constructed from 72.4 million private one-to-one cell phone calls among $3.4$ million individuals in an undisclosed European country over a one-month period \cite{jpweakties,jpweakties2,gonzalez}. This allows the comprehensive ascertainment of ties between people who are customers of the given cell phone operator, and results in a fairly realistic human social network. We keep only reciprocated ties, and denote the number of calls placed between nodes $i$ and $j$ with $w_{ij} = w_{ji}$, which can be conceptualized as tie strength.

Instead of dealing with the entire network, we wish to use several non-overlapping samples of the network with varying properties (size, density, etc.) by slicing it where it most naturally breaks into pieces, which is across communities. To that end, we identify the largest 80  communities \cite{fortunato2009,comnotices,multislice, lehmann2010,weighted} and use them as our samples. To avoid confusion with subsequent node and tie sampling, we refer to these network samples as \emph{community networks}. We detect network communities using modularity maximization in its original formulation \cite{newmangirvan,newmanpnas2006}. Modularity, which is a number lying between -1 and 1, measures how well a given partition $\{c_1, c_2, \ldots, c_N \}$ of a network compartmentalizes its communities, is given by 
\be
Q = \frac{1}{2L}\sum_{i,j} \left[ A_{ij} - \frac{k_i k_j}{2L} \right] \delta(c_i, c_j), 
\ee
where the adjacency matrix element $A_{ij}$ denotes the strength of the tie connecting nodes $i$ and $j$, $k_i$ is the degree of node $i$, $L$ the total weight of the edges (or number of unweighted edges) in the network, $c_i$ the community assignment of node $i$, and $\delta(c_i, c_j)$ is the Kronecker delta function, which is unity if and only if $c_i = c_j$, otherwise it is zero. Modularity, in its original formulation, measures the difference between the total fraction of edges that fall within groups versus the fraction one would expect by chance. The common null model, codified by the $k_ik_j/(2L)$ term, takes degree heterogeneity into account by preserving the expected degree distribution. High values of $Q$ indicate network partitions in which more of the edges fall within groups than expected by chance. While maximizing modularity is known to be an NP-hard problem \cite{BRANDES_2006}, there are numerous computational heuristics available \cite{fortunato2009,comnotices}, and our choice is the so-called Louvain method \cite{BLOND_ARX_2008}.

\begin{figure}
\begin{center}
\includegraphics[width=1\linewidth]{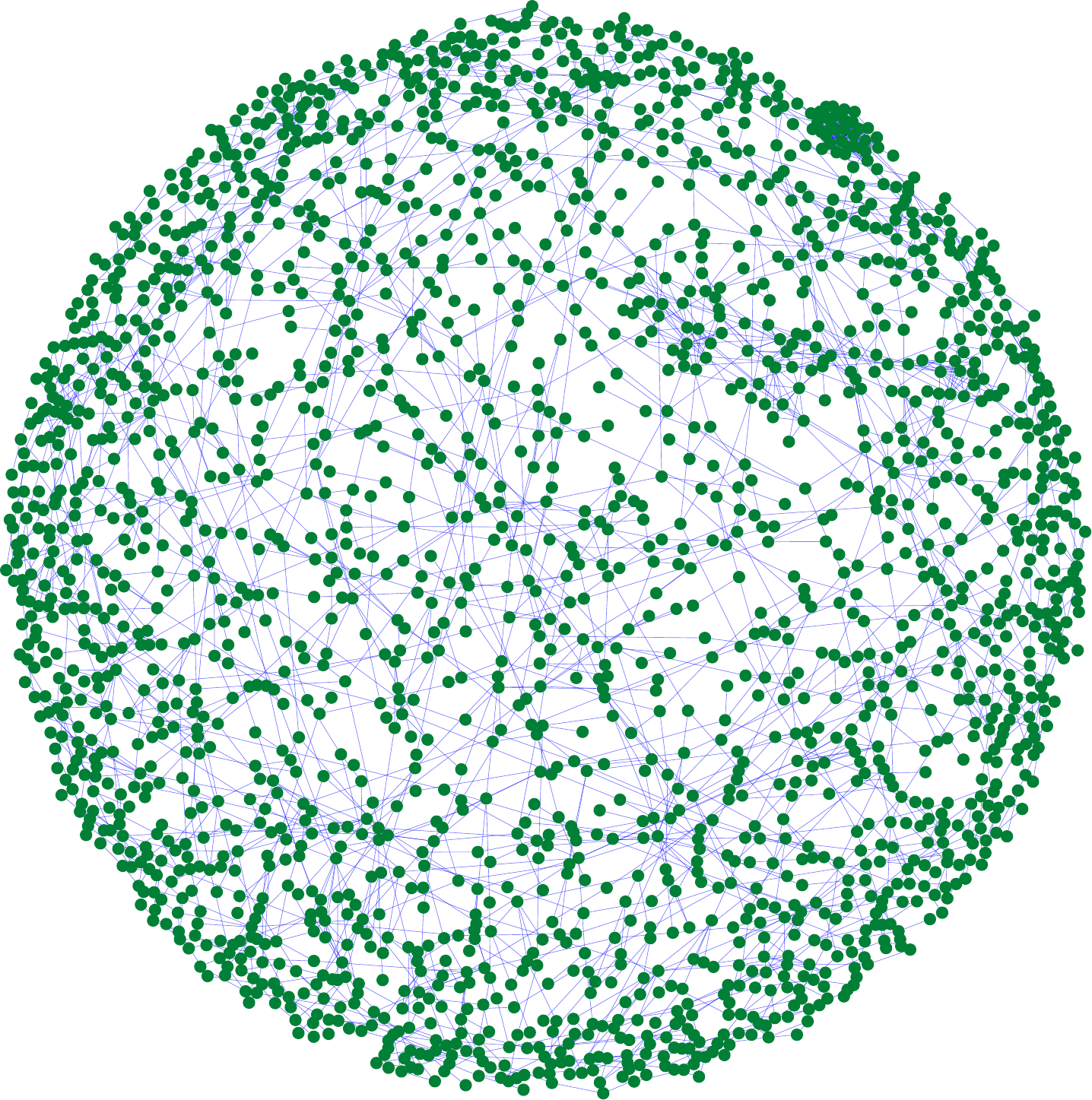}
\caption{\label{fig:egcomm}
Visualization of one of the 80 community networks used in this study. The network consists of one-to-one cell phone calls, and this particular network contains 2130 nodes.}
\end{center}
\end{figure}

\section{Spreading on and sampling of community networks} \label{sec:methods}
Here, we describe the spreading and sampling processes which are carried out on each of the 80 static community networks. We use the canonical Susceptible-Infectious (SI) model, in which each node occupies one of the two states (S or I) \cite{may1992}. The stylized spreading process is carried out in the fully observed community networks, and it proceeds as follows. For each community network, starting from one initially infected seed node, each infected node, per time step, attempts to infect one of its neighbors chosen uniformly at random. The length of a time step is therefore defined as the shortest possible time during which the infection can spread from an infectious node to a susceptible node. For node $j$ with degree $k_j$, this \emph{selection probability} is given by $p_{j} = 1/k_j$, which corresponds to an isotropic one-step random walk. We call this the \emph{unweighted selection} because the choice of the neighbor is topological only, meaning that the neighbor is selected uniformly at random. In contrast, we also use \emph{weighted selection}, where a neighbor $k$ of node $j$ is chosen with probability $p_{jk} = w_{jk} / \sum_m w_{jm}$, where $w_{jm}$ represents the strength of the tie between nodes $j$ and $m$, quantified in terms of call volume as described above, leading to neighbor selection that is biased towards stronger ties.

Once the neighbor has been chosen, the infection happens with \emph{infection probability}, which we have fixed at $0.05$. We run each realization of the simulation for 200 time steps, which is a sufficiently long time, given the value of infection probability, to allow for even very long paths (of the order network diameter) to emerge. We keep track of every infection path by tabulating the predecessors (parents) of each newly infected node, and in case of repeat infections, i.e, an already infected node is made infected for the second time, we only keep track of the first infection event (hence ignoring complex contagion processes \cite{centola2007,centola2010,christakis2010}).

We are interested in the length of the dynamic path $\l$ taken by the infection from the seed node to multiple target nodes. In particular, we now wish to make inferences about path lengths, taken by the spreading process described above, under partial network observation. The latter is achieved using a computational approach, which simulates a twofold ego-centric sampling design. The simulated sampling design is termed conventional because, unlike an adaptive design, it does not use information collected during the ``survey,'' or earlier stages of the sampling process, to direct subsequent sampling \cite{handcock2010}. 

The two stages making up the partial observation are node sampling and tie sampling. First, \emph{node sampling}, for which the units of sampling are nodes, refers to the process of observing a fraction of the nodes, where, moreover, only ties that fall between the observed nodes are retained in the sample. Node sampling, sometimes also called node filtering, therefore corresponds to the idea of observing only a subset of the nodes. We use $f_n$ to denote the fraction of unobserved nodes, such that $1 - f_n$ is the fraction of observed or sampled nodes. The idea of node sampling is similar to the study of random breakdowns of networks in the context of percolation theory. Starting with an initial degree distribution $P(k_0)$, the probability that a node of degree $k_0$ becomes a node of degree $k$, where $k \le k_0$, is given by $\binom{k_0}{k}(1-f_n)^k f_n^{k_0-k}$, and the new degree distribution \cite{cohen2000} becomes
\be
P'(k) = \sum_{k_0=k}^\infty P(k_0)\binom{k_0}{k}(1-f_n)^k f_n^{k_0-k},
\end {equation}
where the post-sampling quantities are denoted by a prime. This leads to an average degree of $\av{k}'=\av{k_0}(1-f_n)$ in the sampled network.

Second, \emph{tie sampling}, for which the units of sampling are network ties, refers to the idea that we typically observe only some fraction of the contacts (neighbors) of each sampled node. It encapsulates the notion that human subjects commonly do not disclose all of their social contacts, a problem that can be partially mitigated by suitable name generators, which are survey instruments used to solicit information from individuals about the people whom they are connected to \cite{campbell1991,marsden2003,marsden1987,barnett2011}. 

For generality, we allow for arbitrary combinations of node and tie sampling. However, when combining the two, we always carry out node sampling first and tie sampling second, which is the order these two processes would occur in a real-world sampling situation. Note that when combining the two sampling processes, the actual number of ties removed in tie sampling is computed from the initial number of ties present in the network prior to node sampling.

To clarify this, consider a network of $N$ nodes and $L$ links. Since the sampled nodes are chosen uniformly at random from the node population, any tie is included in the sample if and only if the adjacent nodes are included. Since each node is included in the sample with probability $(1 - f_n)$, on average a fraction $(1-f_n)^2$ of ties in the network will be included in the sample after \emph{node} sampling. For example, if $f_n = 0.2$, the expected number of ties is $0.64L$. If we subsequently apply tie sampling using, say, $f_e = 0.2$, the expected fraction of ties falls further to $0.64L - 0.2L = 0.44L$. In other words, using these sampling parameters, less than half of the ties in the network would be present in the sample. In general, as a consequence of the full (node \& tie) sampling process, the expected number of nodes in the sample is $N' = (1-f_n)N$, whereas the expected number of ties in the sample is $L' =  [(1-f_n)^2 - f_e]L$.

\section{Simulation results} \label{sec:results}
In this Section, we report results on three different types of inference. First, to what extent do path lengths $\ls$ in a partially observed or sampled network represent path lengths $\lu$ in the underlying unsampled network? Second, if it were possible to observe the network fully, how well would topological paths represent the actual (unobserved) dynamic paths as followed by the process? Third, if the network were to be only partially observed, how well do sampled topological paths represent the actual (unobserved) dynamic paths followed by the process? Note that the first question is strictly topological, while the second and third questions are affected by both network topology and process dynamics.

To quantify these biases, we define three bias factors, where the averages are taken over different process realizations. First, the ratio of sampled path length to unsampled path length as a function of actual path length is denoted
\be
b_1(\l) = \frac{\la \ls(\l) \ra}{\la \lu(\l) \ra} \ge 1
\ee
since $\ls(\l) \ge \lu(\l)$ for all $\l$; second, the ratio of unsampled path length to the actual path length is
\be
b_2(\l) = \frac{\la \lu(\l) \ra}{\l} \le 1
\ee
since $\l \ge \lu(\l)$ for all $\l$; and, third, the ratio of sampled path length to the actual path length
\be
b_3(\l) = \frac{\la \ls(\l) \ra}{\l} > 0
\ee
but is otherwise unbounded. The corresponding averages are
\be
b_1 = \left \langle b_1(\l) \right \rangle, \,\, b_2 = \left \langle b_2(\l) \right \rangle, \,\, b_3 = \left \langle b_3(\l) \right \rangle,
\ee
where the averages are taken over a range of values for $\l$. For any network, $b_1 \ge 1$, $b_2 \le 1$, and $b_3 > 0$. 

\begin{figure*}
\begin{center}
\includegraphics[width=0.9\linewidth]{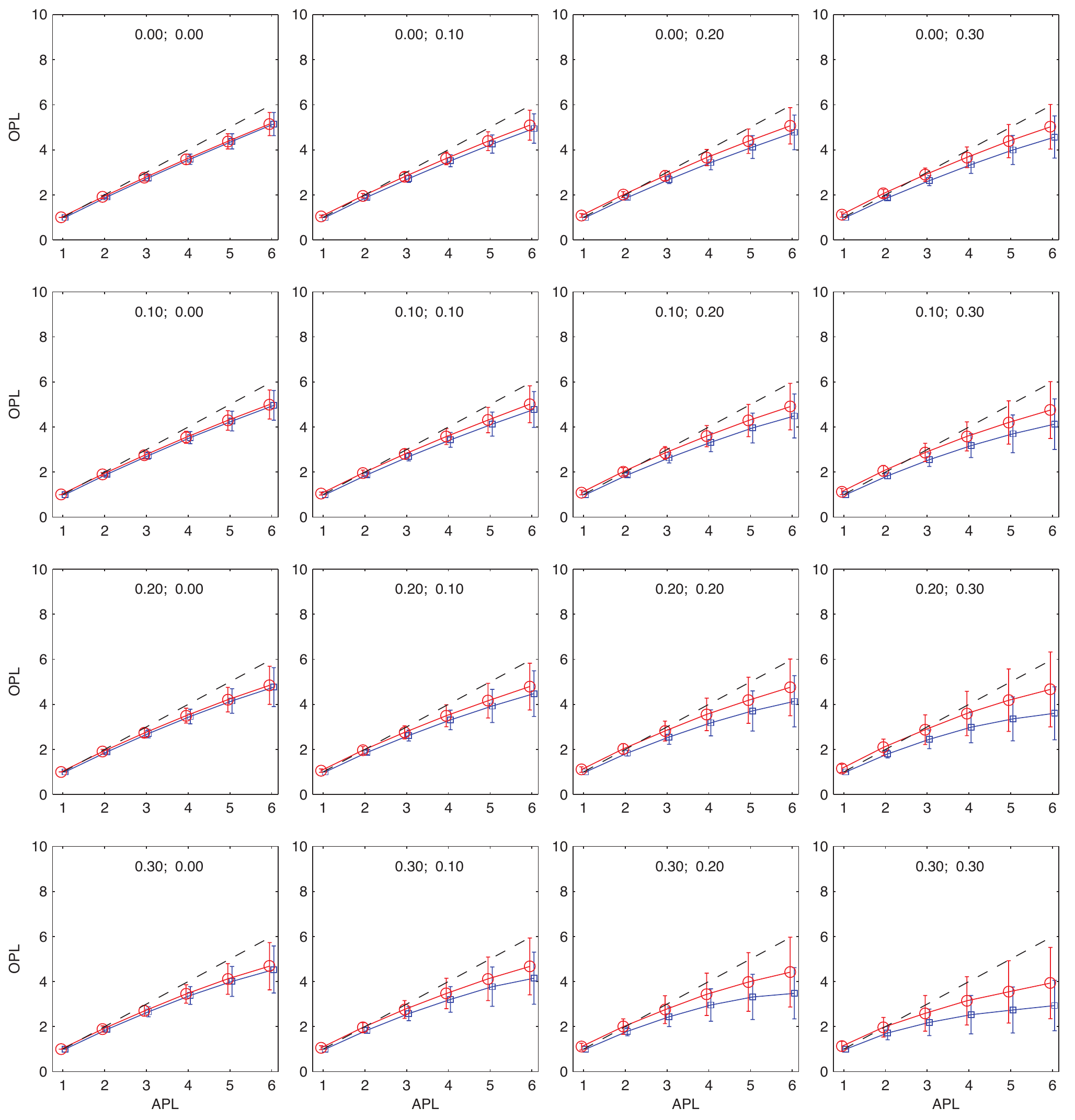}
\caption{\label{fig:paths1}
(Color online)  
Observed path lengths (OPL) as a function of the actual path lengths (APL) in a medium-size community of $N \approx 2,000$ nodes. Each panel corresponds to a different fraction of unobserved nodes and unobserved ties as indicated in each panel by the $(f_n, f_e)$ pair. The blue curves correspond to the average unsampled path lengths $\lu$ and the red curves to the average sampled path lengths $\ls$. The extent of fluctuations is indicated with the error bars, which are given as  $\lu(\l) \pm \sigma_u(\l)$ for unsampled paths and $\ls(\l) \pm \sigma_s(\l)$ for sampled paths. The diagonal dashed black lines corresponds to the identity relationship, i.e., points where the observed path lengths are identical to the actual path lengths. For this particular community, the sampled path lengths are typically shorter than the actual path lengths.}
\end{center}
\end{figure*}

\begin{figure*}
\begin{center}
\includegraphics[width=0.9\linewidth]{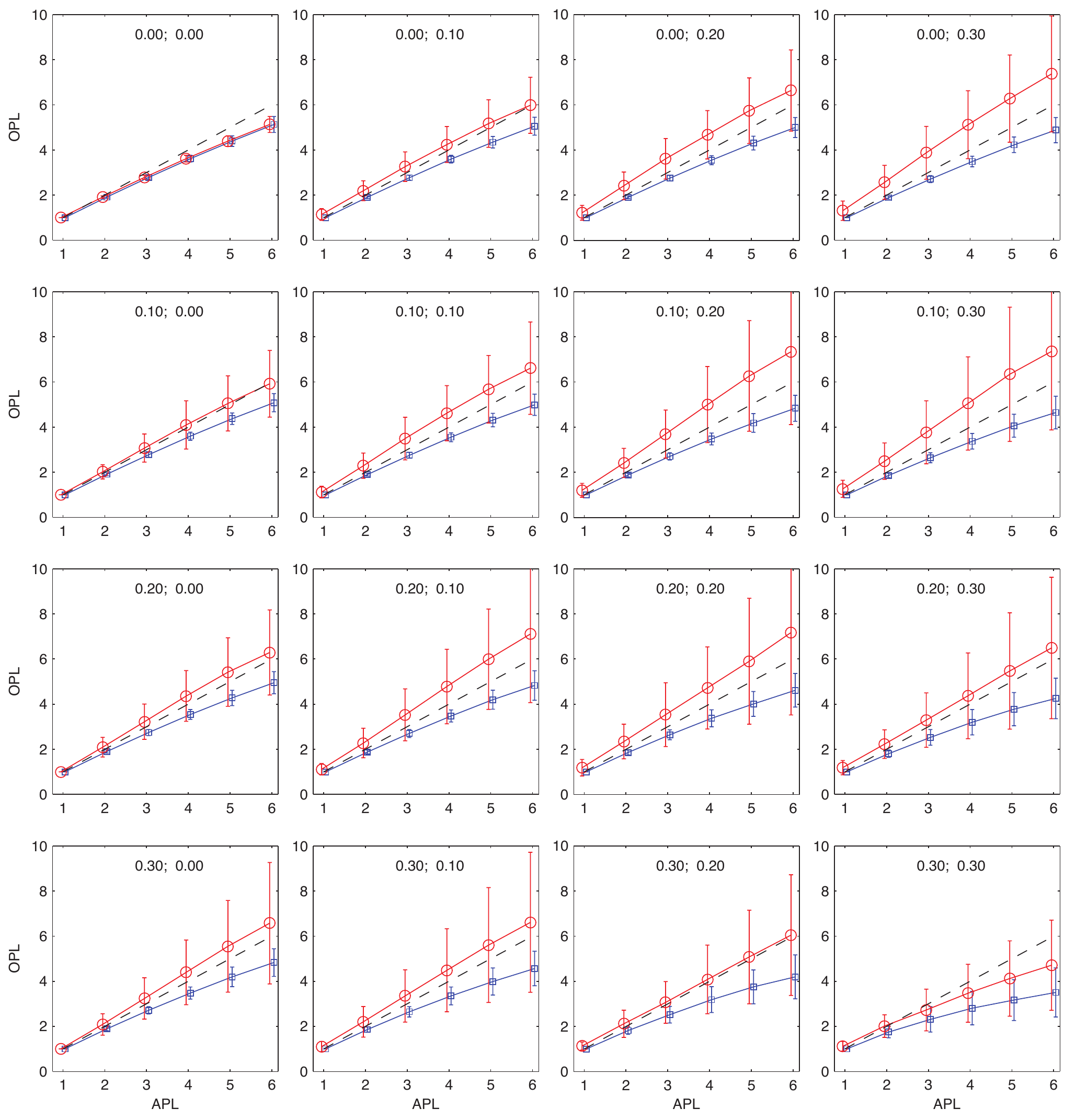}
\caption{\label{fig:paths2}  (Color online)
Observed path lengths (OPL) as a function of the actual path lengths (APL) in a large community of $N \approx 20,000$ nodes. For this particular community, the average sampled path lengths $\ls$ may be longer or shorter than the actual path lengths, depending on the values of $f_n$ and $f_e$.}
\end{center}
\end{figure*}

In Figs.~\ref{fig:paths1} and \ref{fig:paths2}, we show the average path lengths using unweighted neighbor selection for community networks of $\sim$2,000 nodes and $\sim$20,000 nodes, respectively, averaged over $1,000$ attemped realizations (see discussion below), for the sampled path lengths $\ls$, shown in red, and the unsampled path lengths $\lu$, shown in blue, as a function of the actual path length $\l$ as followed by the infection process. Since the process is run 1,000 times for each combination of sampling parameter values $(f_n, f_e)$, each dot represents an average. To quantify the extent of fluctuations around the average, we also compute standard deviations, such that the plotted function can be expressed as $\lu(\l) \pm \sigma_u(\l)$ for unsampled paths and $\ls(\l) \pm \sigma_s(\l)$  for sampled paths, where $\sigma_u(\l)$ and $\sigma_s(\l)$ are the corresponding standard deviations. Note that the sampled path lengths $\ls$ are necessarily as long as or longer than the unsampled path lengths $\lu$, meaning that the red curve always lies on or above the blue curve. The distance between the red and blue curves describes the bias due to approximating the (unobserved) shortest paths in the original network with the (sampled) paths in the perturbed network and is quantified by $b_1$. Note also that the blue curve is always on or below the black line, consistent with the fact that the actual path can never be shorter than the shortest path (by definition). The distance between the blue curve and the black line is the bias due to not having observed the spreading process, but instead approximating it with shortest paths computed in the (typically unobserved) original network. This bias is quantified by $b_2$. Finally, the gap between the red and the black line is the bias due to not having observed the spreading process but, instead, approximating it with sampled shortest paths, i.e., shortest paths computed in the perturbed network. The extent of this bias is quantified by $b_3$.

\begin{figure*}
\begin{center}
\includegraphics[width=0.9\linewidth]{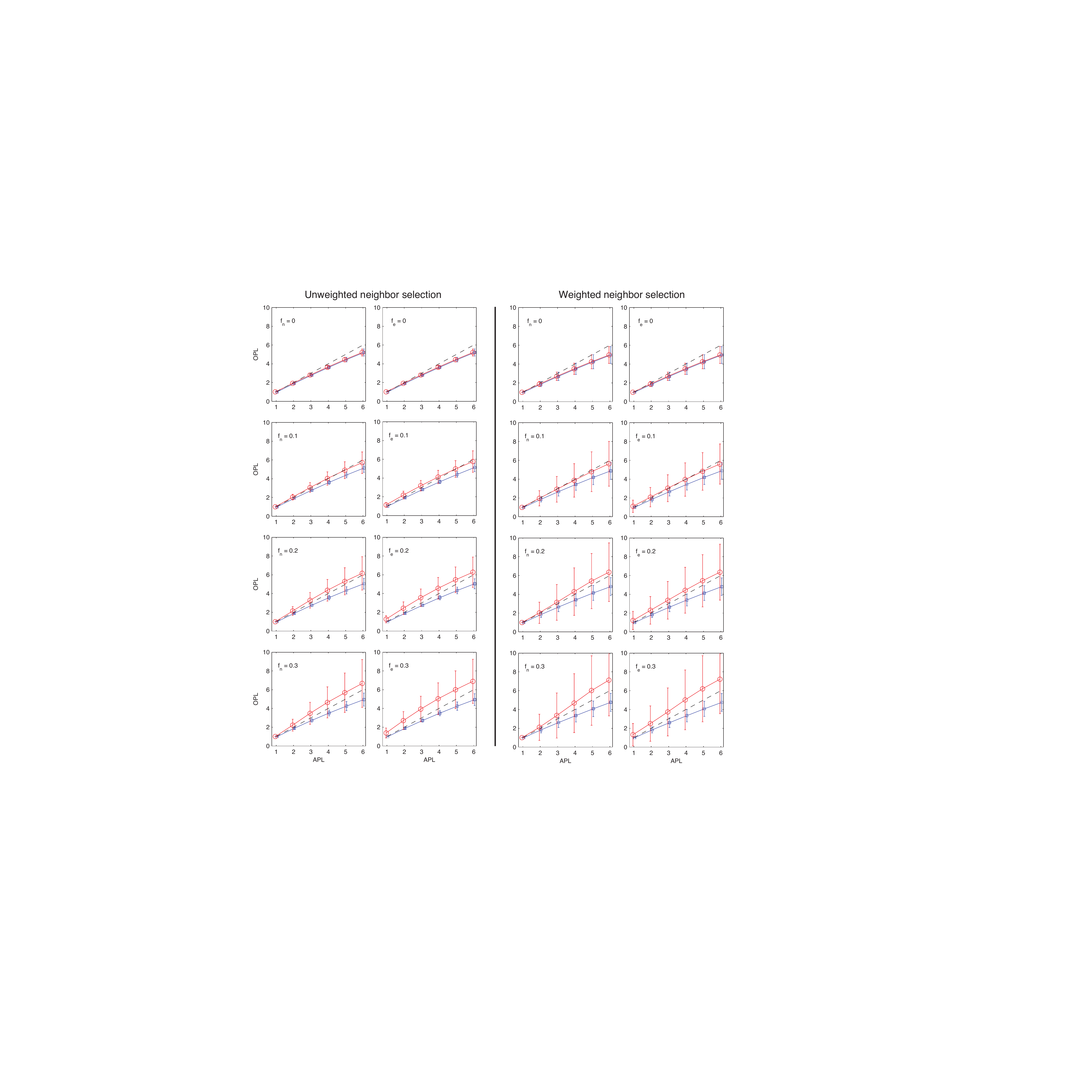}
\caption{\label{fig:paths3uw} 
(Color online) Observed path lengths (OPL) as a function of the actual path lengths (APL) averaged over 80 communities using unweighted neighbor selection (8 panels on the left) and weighted neighbor selection (8 panels on the right). For each type of neighbor selection, the leftmost column corresponds to node sampling only, where the value of $f_n$ is indicated in the panel, and $f_e = 0$ for all panels (i.e. there is no edge sampling). The rightmost columns for each type of neighbor selection correspond to edge sampling, where the value of $f_e$ is indicated in the panel, and $f_n = 0$ for all panels (i.e. there is no node sampling). Shortest paths in partially observed networks typically overestimate the actual path lengths, the extent of which depends on the sampling parameters as well as the length of the actual path $\l$ taken by the process. Note that weighted neighbor selection in the spreading process introduces considerable fluctuations, meaning that if the process is sensitive to tie strengths, sampled topological paths reflect the actual process path length poorly, and may either significantly overestimate or underestimate the path length.}
\end{center}
\end{figure*}

Depending on the network, the sampled paths may be longer or shorter than the actual paths, but which of these outcomes is more typical? For any of the 80 community networks, and for any of the 49 unique $(f_n, f_e)$ sampling parameter combinations, and for any value of the the  actual path length $\{1,2,3,4,5,6\}$, we obtain $1,000$ attempted realizations for $\lu$ and $\ls$. We say attempted because the more we sample, the thinner the resulting network becomes, and consequently the smaller the number of paths of any given length in the sampled network. Under heavy sampling, it is possible that not every realization contains of path of length, say, $\l=6$. For this reason, when computing the mean path lengths and the standard deviations, the statistics need to be weighted. To accomplish this, let us first expand our earlier notation slightly. We let $\lu(\l, \eta)$ and $\ls(\l, \eta)$ represent the average path lengths, unsampled and sampled, respectively, at distance $\l$ for network $\eta$; similarly we let $\sigma_u(\l, \eta)$ and $\sigma_s(\l, \eta)$ represent the corresponding standard deviations of the path lengths; and finally $n_u(\l, \eta)$ and $n_s(\l, \eta)$ are the number of observations in each category, which are less than or equal to $10^3$, the number of attempted realizations in each category. The values of $f_n$ and $f_e$ are considered fixed. The ensemble mean for sampled paths is now given by
\begin{equation}
\la \ls(\l) \ra_{\eta} = \frac{1}{\sum_{\eta = 1}^{n} n_s(\l, \eta)} \sum_{i=1}^{n} n_s(\l, i) \ls(\l, i),
\end{equation}
and the ensemble standard deviation is given by 
\begin{widetext}
\begin{equation}
\la \sigma_s(\l) \ra_{\eta} = \sqrt{\frac{1}{\sum_{\eta = 1}^{n} n_s(\l, \eta)}\sum_{i=1}^{n} n_s(\l, i) \sigma^2_s(\l, i)  + \frac{1}{\left( \sum_{\eta = 1}^{n} n_s(\l, \eta) \right)^2}\sum_{i=1}^{n}\sum_{j=i+1}^{n} n_s(\l, i) n_s(\l, j) \big[\ls(\l, i) - \ls(\l, j) \big]^{2}},
\end{equation}
\end{widetext}
where $\la \ls(\l) \ra_{\eta}$ is simply a weighted mean of the means, whereas $\la \sigma_s(\l) \ra_{\eta}^2$ has two components, the former being a weighted mean of the variances, and the latter being a weighted mean of the squares of all pairwise differences of the means. Both results  follow from a direct calculation, and the expressions for the unsampled paths are identical and follow by changing the subscripts from $s$ to $u$. We show the plots of $\la \lu(\l) \ra_{\eta} \pm \la \sigma_u(\l) \ra_{\eta}$ and $\la \ls(\l) \ra_{\eta} \pm \la \sigma_s(\l) \ra_{\eta}$ for both unweighted and weighted neighbor selection in Fig.~\ref{fig:paths3uw} As expected, the average unsampled paths underestimates the actual path lengths, and the extent of this bias increases as $\l$ increases. The sampled path lengths may however overestimate or underestimate the actual path lengths. While the averages behave very similarly, there are significant differences in fluctuations between the unweighted and weighted spreading process. While the weighted process in general shows more fluctuations, the extent of fluctuations is especially pronounced for sampled path lengths. In other words, the weighted spreading process may veer the dynamic path even further from the structurally shortest paths.

\begin{figure}
\begin{center}
\includegraphics[width=0.80\linewidth]{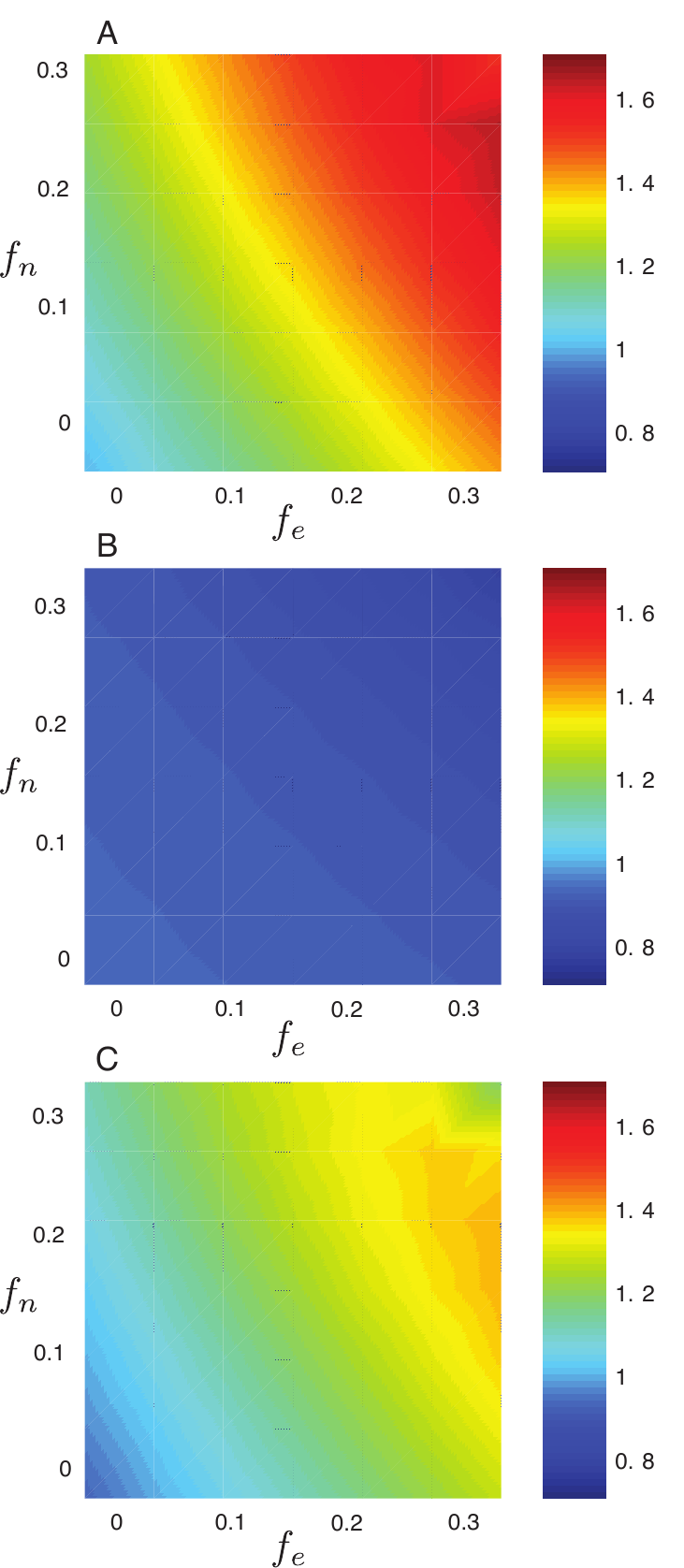}
\caption{\label{fig:ave}  (Color online)  
Surface plots of sampling bias as a function of $f_e$, the fraction of removed edges (horizontal axes) and $f_n$, the fraction of removed nodes (vertical axes) for the unweighted neighbor selection as described in Section \ref{sec:methods}.
\textbf{(a)} Plot of $b_1$, the average ratio of sampled path lengths to unsampled path lengths. 
\textbf{(b)} Plot of $b_2$, the average ratio of unsampled path lengths to to actual path lengths.
\textbf{(c)} Plot of $b_3$, the average ratio of sampled path lengths to to actual path lengths.
The results are essentially identical for the weighted neighbor selection.}
\end{center}
\end{figure}

The average outcomes are surprisingly similar for unweighted and weighted neighbor selection, which could have its origin in how the community networks are sampled and the connection between network structure and tie strength as quantified by the weak ties hypothesis \cite{granovetter,jpweakties}. To elaborate on this, we would expect community networks to have a high density of ties, higher than what would be expected by chance. The weak ties hypothesis, on the other hand, states that there is a positive association between the fraction of shared friends any two connected individuals $i$ and $j$ have and the strength of the tie $w_{ij}$ connecting them. This suggests that most ties within communities would be expected to be fairly strong and, consequently, the impact of incorporating weights in the neighbor selection process might be fairly small. However, as indicated above, the extent of fluctuations is much greater for the weighted neighbor selection than for the unweighted one.

In order to express the bias for all examined path lengths $\l = 1, \ldots, 6$, and over all 80 community networks, we computed the conditional averages $\langle b_1 | f_n, f_e \rangle$, $\langle b_2 | f_n, f_e \rangle$, and $\langle b_3 | f_n, f_e \rangle$, which quantify the overall bias for given levels of node and tie sampling, and they are shown in Fig.~\ref{fig:ave}. The underlying numerical values are given in Table I. For example, using $f_n = f_e = 0.2$, which implies that after sampling 44\% of ties remain in the network, results in $\langle b_1 | f_n = 0.20, f_e = 0.20\rangle = 1.50$, showing that sampled paths are 50\% longer than unsampled paths for the given level of node and tie sampling; $\langle b_2 | f_n = 0.20, f_e = 0.20\rangle = 0.88$ shows that the unsampled topological paths are 88\% of the length of the stochastic paths; and finally $\langle b_3 | f_n = 0.20, f_e = 0.20\rangle = 1.32$ shows that sampled topological paths over-estimate path length by 32\%.

\begin{figure}
	\centering
        \includegraphics[width=1\linewidth]{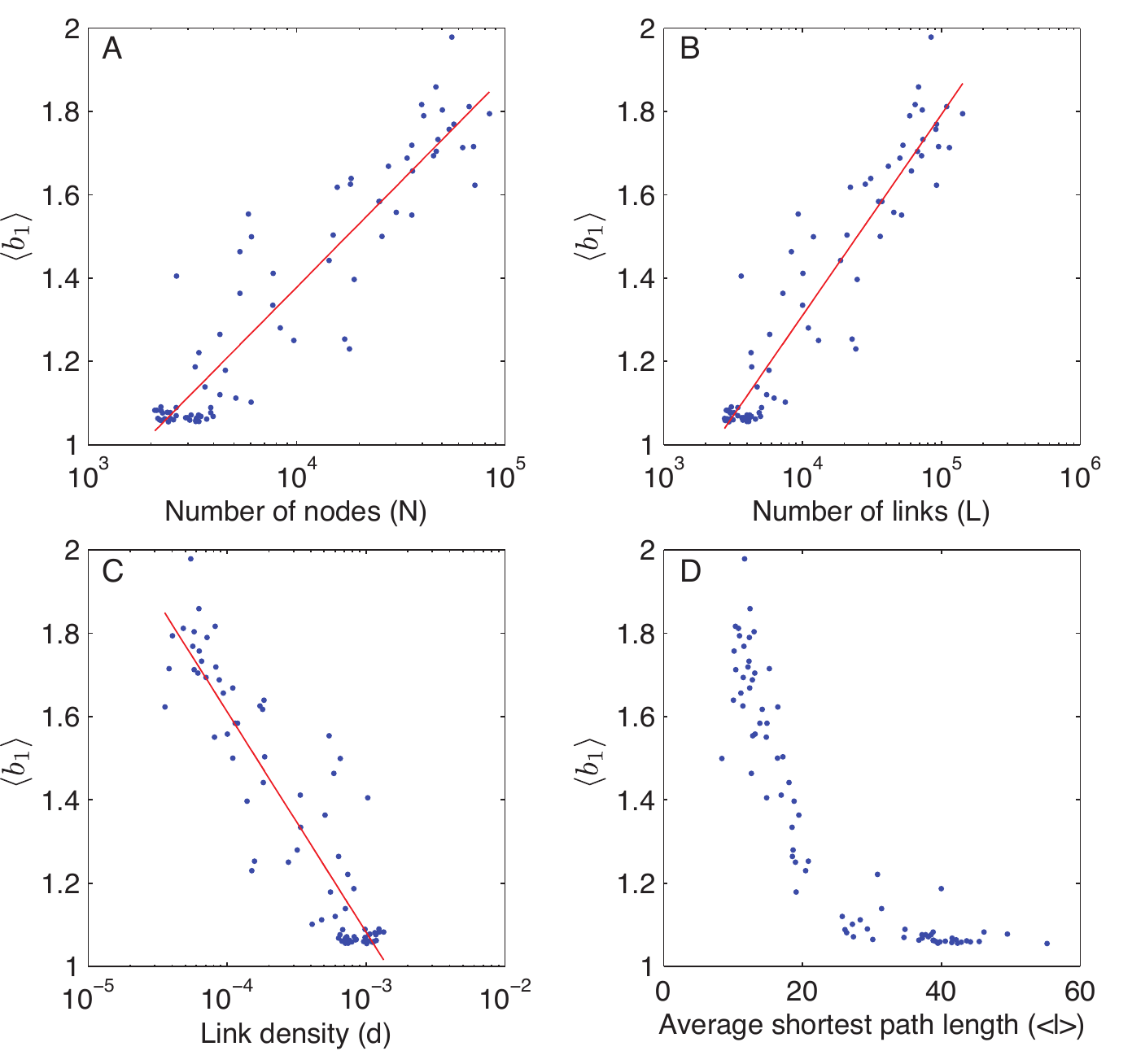}
\caption{(Color online) Average bias-ratios $\langle b_1 \rangle$ (sampled path length divided by unsampled path length) as a function of different network characteristics for the studied 80 subnetworks. The value of $\langle b_1 \rangle$ is always greater than one.}
\label{fig:sum1}
\end{figure}

\begin{figure}
	\centering
        \includegraphics[width=1\linewidth]{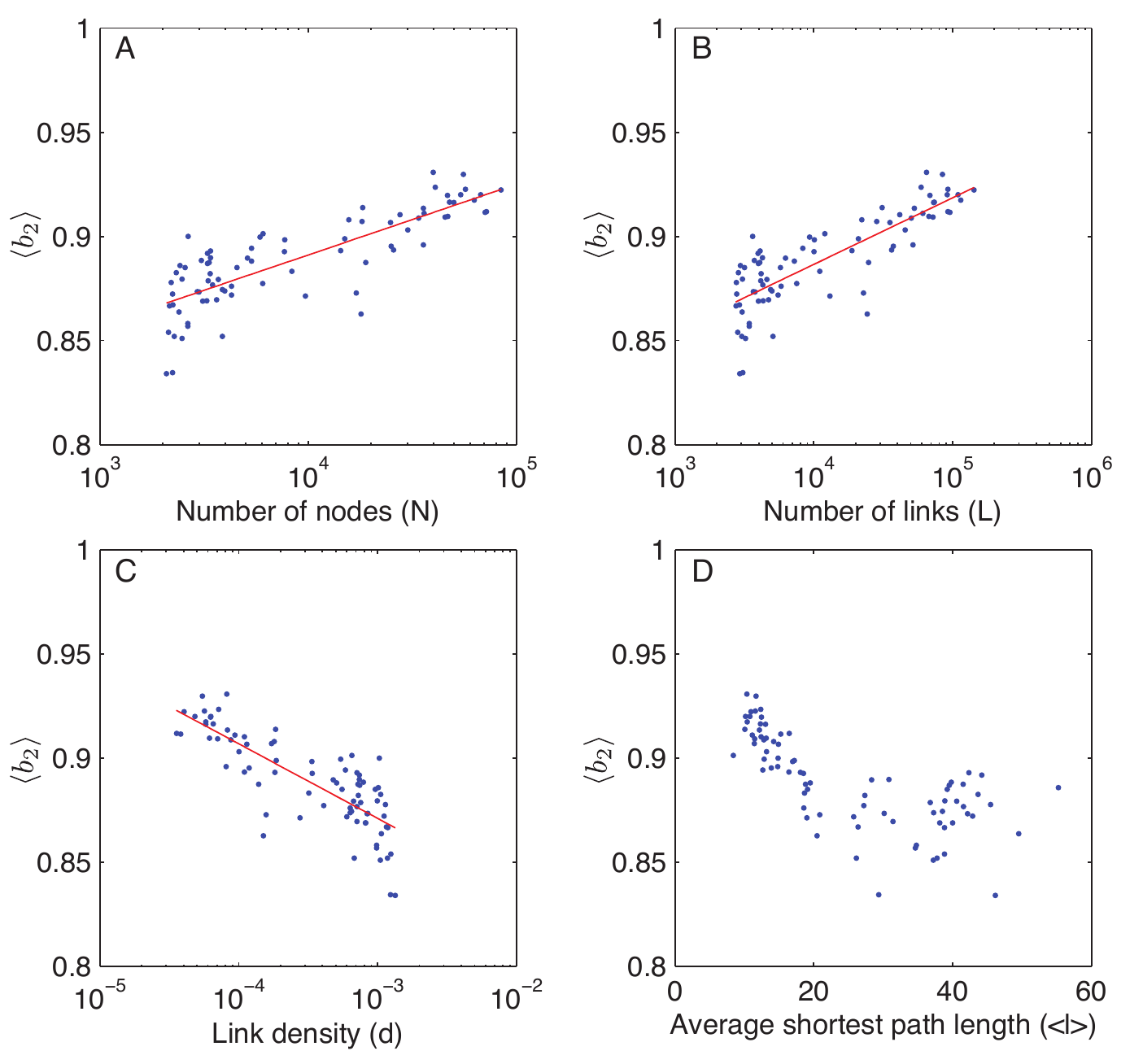}
\caption{(Color online) Average bias-ratios $\langle b_2 \rangle$ (unsampled path length divided by dynamic path length) as a function of different network characteristics. The value of $\langle b_2 \rangle$ is always less than one.}
\label{fig:sum2}
\end{figure}

\begin{figure}
	\centering
        \includegraphics[width=1\linewidth]{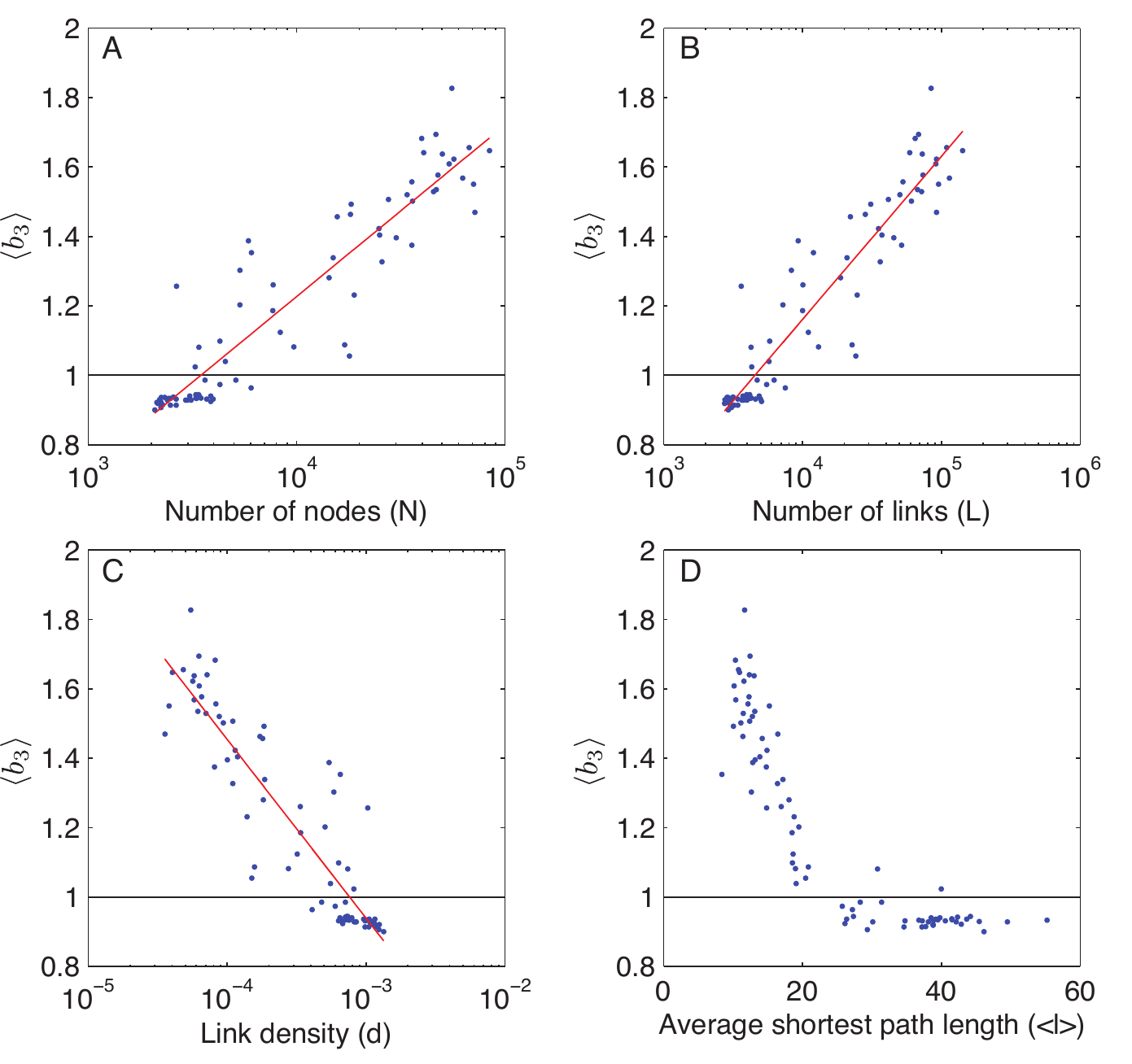}
\caption{(Color online) Average bias-ratios $\langle b_3 \rangle$ (sampled path length divided by dynamic path length) as a function of different network characteristics. The value of $\langle b_3 \rangle$ may be less than one or greater than one, depending on the community network. The horizontal line corresponds to $\langle b_3 \rangle = 1$.}
\label{fig:sum3}
\end{figure}

The above averages, although informative, mask the variation from one community network to another. Therefore, instead of averaging over community networks, we average, each network, over the sampling parameters $f_n$ and $f_e$. Figs.~\ref{fig:sum1}, \ref{fig:sum2}, and \ref{fig:sum3} show the value of this average bias plotted against network size $N$, number of links $L$, link density $d = 2L / N(N-1)$, and average shortest path length $\langle \l \rangle$ for all 80 subnetworks. To three of the four plots in each figure, we fitted a linear regression model of the form $\langle b \rangle = \beta_0 + \beta_1 \log(x)$, where $x$ is either $N$, $L$, or $d$. To gauge the goodness of fit of the model, we used the simple (non-adjusted) $R^2$ statistic. For each bias factor, $\langle b_1 \rangle$, $\langle b_2 \rangle$, and $\langle b_3 \rangle$, we find that most variance is always explained by $L$ (number of links), then by $N$ (number of nodes), and finally by $d$ (link density), although the three typically come close to one another. The $R^2$ values using $L$ as the predictor of bias are $0.89$, $0.63$, and $0.88$ for  $\langle b_1 \rangle$, $\langle b_2 \rangle$, and $\langle b_3 \rangle$, respectively, and the corresponding parameter values of interest are $\beta_1 = 0.4826$ (for $\langle b_1 \rangle$), $\beta_1 = 0.0320$ (for $\langle b_2 \rangle$), and $\beta_1 = 0.4698$ (for $\langle b_3 \rangle$). Consequently, of the three, the value of $\langle b_2 \rangle$ is by far the least sensitive to variation in $L$ (or $N$ or $d$; see Fig.~\ref{fig:sum2}).

Therefore, using unsampled topological paths for stochastic paths typically results in a fairly small overall bias, and the bias is always downwards as expected, and therefore the resulting values for $\langle b_2 \rangle$ are always below one. In contrast, using sampled topological paths for stochastic paths may result in an upward or downward bias, depending the network and the sampling parameters, such that $\langle b_3 \rangle$ may be less than one or more than one. The extent of this bias is well predicted by the number of links $L$ in the network, and the value $\beta_1 = 0.4698$ suggests that multiplying the number of links by a factor of ten results in an addition of $0.47$ in its value. Of the studied 80 community networks, 35 had $\langle b_3 \rangle$ less than one; based on the results of the regression models, in particular the locations where the regression lines meet the (horizontal) no-bias lines, these networks have typically less than 3,500 nodes, less than 5,000 links, high link density ($d > 0.0008$), and average shortest path length greater than 25. In other words, compared to the population of studied community networks, these tend to be small and relatively densely connected networks.

\begin{table}
\begin{center}
\begin{tabular}{c|ccccccc}
$f_n | f_e$& 0.00 & 0.05 & 0.10 & 0.15 & 0.20 & 0.25 & 0.30\\
\hline
0.00 &1.00 &1.07 &1.14 &1.20 &1.27 &1.34 &1.41\\
0.05 &1.05 &1.11 &1.18 &1.25 &1.33 &1.40 &1.48\\
0.10 &1.08 &1.16 &1.23 &1.31 &1.38 &1.46 &1.54\\
0.15 &1.12 &1.20 &1.28 &1.36 &1.45 &1.52 &1.59\\
0.20 &1.16 &1.24 &1.33 &1.42 &1.50 &1.57 &1.63\\
0.25 &1.20 &1.29 &1.38 &1.46 &1.54 &1.60 &1.64\\
0.30  &1.23 &1.33 &1.42 &1.50 &1.57 &1.60 &1.51\\
\end{tabular}
\end{center}

\begin{center}
\begin{tabular}{c|ccccccc}
$f_n | f_e$& 0.00 & 0.05 & 0.10 & 0.15 & 0.20 & 0.25 & 0.30\\
\hline
0.00 &0.93 &0.92 &0.92 &0.91 &0.91 &0.91 &0.90\\
0.05  &0.92 &0.92 &0.91 &0.91 &0.90 &0.90 &0.89\\
0.10 &0.92 &0.92 &0.91 &0.90 &0.90 &0.89 &0.88\\
0.15 &0.92 &0.91 &0.90 &0.90 &0.89 &0.88 &0.86\\
0.20 &0.91 &0.90 &0.90 &0.89 &0.88 &0.86 &0.84\\
0.25 &0.91 &0.90 &0.89 &0.88 &0.86 &0.84 &0.82\\
0.30 &0.90 &0.89 &0.88 &0.86 &0.84 &0.81 &0.77\\
\end{tabular}
\end{center}

\begin{center}
\begin{tabular}{c|ccccccc}
$f_n | f_e$& 0.00 & 0.05 & 0.10 & 0.15 & 0.20 & 0.25 & 0.30\\
\hline
0.00 &0.93 &0.99 &1.05 &1.10 &1.16 &1.22 &1.27\\
0.05 &0.96 &1.02 &1.08 &1.14 &1.20 &1.26 &1.32\\
0.10 &1.00 &1.06 &1.12 &1.18 &1.25 &1.31 &1.36\\
0.15 &1.03 &1.09 &1.16 &1.22 &1.29 &1.34 &1.39\\
0.20 &1.06 &1.12 &1.19 &1.26 &1.32 &1.36 &1.39\\
0.25 &1.08 &1.16 &1.23 &1.29 &1.33 &1.36 &1.35\\
0.30 &1.11 &1.18 &1.25 &1.30 &1.33 &1.31 &1.17\\
\end{tabular}
\end{center}
\label{tab3}
\caption{Values of different bias ratios. Top panel: $\langle b_1 | f_n, f_e \rangle$, the average ratio of sampled shortest path lengths to unsampled shortest path lengths. Middle panel: $\langle b_2 | f_n, f_e \rangle$, the average ratio of unsampled shortest path lengths to actual path lengths. Bottom panel: $\langle b_3 | f_n, f_e \rangle$, the average ratio of sampled shortest path lengths to actual path lengths. All ratios are tabulated according to $f_n$ and $f_e$. The value in the bottom right corner for $f_n = f_e = 0.30$ in the top and bottom panels deviates from the trend present in the two tables. As this value corresponding to the greatest degree of sampling (both tables deal with sampled path lengths) and hence to the least number of data points in the average, it is likely a statistical fluctuation.} 
\end{table}

\section{Discussion} \label{sec:discussion}
{The last few years have seen a strong emphasis in the literature on understanding structural properties of complex networks, although increasingly the field appears to be moving in the direction of network dynamics, where dynamics can be understood both as dynamics of networks and dynamics on networks. Spreading and diffusion processes are the archetypes of dynamical processes on networks. In this paper, we have explored the connection between structural (topological) shortest paths, which are elementary network characteristics and on which others measures, such as betweenness centrality, are based, and the lengths of certain types of functional (dynamic) spreading paths. We have introduced the additional layer of network sampling which is relevant from an empirical point of view but which, as we have seen, typically complicates the relationship between structural and functional paths.

More specifically, we have considered the properties of three different types of paths in social networks. In particular, we have compared their lengths under partial network observation, i.e. when there is sampling at the level of nodes, ties, or both. The paths we studied were: (1) the stochastic path taken by a spreading process from source to target, which is known in simulations; (2) the shortest path from source to target in a fully observed network; and (3) the shortest path from source to target in a partially observed network.

Our findings counteract the naive intuition that sampling will always inflate path lengths, in other words, the notion that dealing with a partially observed network would necessarily make processes seem to travel farther than the actually do. The shortest path between any two nodes in a partially observed network will, of course, be as long or longer than the shortest path between the same nodes in a fully observed network. However, in some cases, the upward bias caused by partial observation, the extent of which depends on the structure of the underlying network, can be offset by the tendency of spreading processes to take non-optimal (longer than shortest) paths, the extent of which depends on the details of the spreading process. In some of the community networks studied, the sampled path lengths were always shorter than the actual path lengths, while in other networks either could be shorter, depending on the extent and nature (nodes vs. ties) of sampling. We found that when averaged over all community networks, there were more fluctuations present for the weighted process than for the unweighted one. In particular, the fluctuations were especially pronounced for sampled paths.

Since social networks are almost never fully observed, even if some facet of them might be, such as electronic communication records under ideal circumstances, it is important to understand the impact of sampling on path lengths, and it is likely to find many applications. For example, in a recent study, in addition to epidemiological and genomic data, Gardy and coauthors used a social network constructed from patient interviews to determine the origin and transmission dynamics of a tuberculosis outbreak \cite{gardy2011}. Traditional contact tracing (the identification and diagnosis of persons who may have come into contact with an infected person) did not identify a probable source. However, the structure of the elicited social network suggested ``the most likely source'' of the epidemic. Although it is not clear how the source was identified, it was likely inferred from (partially observed) topological shortest paths.

Another recent study by Rocha, Liljeros, and Holme studied a network of alleged offline sexual contacts between anonymous escorts and sex buyers as self-reported by both parties in an online  community \cite{rocha2010}. Approximately 71\% of the individuals in the largest connected component were reachable by following the time-ordering of the contacts, suggesting that a majority of the component was connected in a way that would allow sexually transmitted diseases to spread between its members \cite{rocha2010}. In this case, time-ordered data were available, which strongly limits the possible spreading paths, given that the contacts need to happen in a certain temporal sequence to potentially transmit a harmful virus or bacterium. Nevertheless, the system is a sample of the underlying population, since the buyers and sellers could be sexually active with individuals not members of the online community. If one were to calculate, for example, how far a given strain of the HIV could have travelled and, hence, how many individuals might have been exposed to it, misestimating the path lengths might lead to misestimates of the size of the epidemic.

There are three obvious ways to extend our work. First, there is the structure of the underlying network, and the results are expected to vary significantly as the topology of this platform is varied. Second, there are the details of the spreading process, which could be modified to be more realistic, and could be tailored towards specific illnesses. Further, to study the spread of behaviors and norms, it might be fruitful to include ideas from the growing literature on complex contagions \cite{centola2007,centola2010,christakis2010}. Third, in our sampling scheme, the units of sampling were either nodes, ties, or both nodes and ties, but one could study the phenomenon for more realistic sampling designs, such as respondent driven sampling (RDS) used to study small but important hard-to-reach populations, such as injection drug users \cite{goel2010}. Finally, although we have framed the problem in the context of social networks, the concepts are generic, and they could be applied to any type of network for which an understanding of the permeation depths of dynamic processes in sampled data are important.

\textbf{Acknowledgements:} We thank A.-L. Barab\'asi for sharing the mobile call data used as part of this research, and J.~Fernandez-Gracia for reading the manuscript.

\providecommand{\noopsort}[1]{}\providecommand{\singleletter}[1]{#1}%

\end{document}